\documentclass[letter,twocolumn]{jpsj2} 

\def\ef{\epsilon_{\rm F}}

\def\ggs{\buildrel\textstyle > \over {\hbox{\raise0.2ex\hbox{$\sim$}}}}
\def\lls{\buildrel\textstyle < \over {\hbox{\raise0.2ex\hbox{$\sim$}}}}
\def\gsim{\,\lower0.75ex\hbox{$\ggs$}\,}
\def\lsim{\,\lower0.75ex\hbox{$\lls$}\,}

\def\tt{\tilde{t}}

\def\runauthor{H. Seo {\it et al.}}

\title{
Single-Component Molecular Metals as Multiband $\pi$-$d$ Systems} 

\author{
Hitoshi \textsc{Seo}\thanks{email address: seo0@spring8.or.jp},
Shoji \textsc{Ishibashi}$^{1}$, 
Yoshinori \textsc{Okano}$^{2}$, 
Hayao \textsc{Kobayashi}$^{3}$,
Akiko \textsc{Kobayashi}$^{3}$, \\
Hidetoshi \textsc{Fukuyama}$^{4}$, 
and 
Kiyoyuki \textsc{Terakura}$^{5}$}
\inst{Synchrotron Radiation Research Center, Japan Atomic Energy Agency, SPring-8, Sayo, Hyogo 679-5148\\
$^{1}$Research Institute for Computational Sciences, \\
National Institute of Advanced Industrial Science and Technology, 
Tsukuba, Ibaraki 305-8568\\
$^{2}$Institute for Molecular Science, Okazaki, Aichi 444-8585\\
$^{3}$Department of Chemistry, College of Humanities and Sciences, Nihon University, Setagaya-ku, Tokyo 156-8550\\
$^{4}$Department of Applied Physics, Faculty of Science, Tokyo University of Science, Shinjuku-ku, Tokyo 162-8601\\
$^{5}$Japan Advanced Institute of Science and Technology, Nomi, Ishikawa 923-1292}

\abst{
Electronic states of single-component molecular metals $M$(tmdt)$_2$ ($M$=Ni, Au) 
are studied theoretically. 
We construct an effective three-band Hubbard model for each material 
by numerical fitting to first-principles band calculations, 
while referring to molecular orbital calculations for the isolated molecules. 
The model consists of two kinds of base orbital for each molecule with hybridization between them, 
 i.e., 
 a $\pi$-character orbital for each of the two tmdt ligands, 
 and, a $pd\pi$-orbital for $M$=Ni or a $pd\sigma$-orbital for $M$=Au centered on the metal site; 
 this indicates that these materials can be considered as novel multiband $\pi$-$d$ systems. 
We find that both orbitals contribute to realize the metallic character in Ni(tmdt)$_2$. 
The origin of the antiferromagnetic transition observed in Au(tmdt)$_2$ 
is also discussed based on this model. 
}

\kword{single-component molecular metals, molecular conductors, 
$\pi$-d system, Hubbard model, 
first-principles band calculation, molecular-orbital, 
magnetism}

\begin{document}
\maketitle

\begin{figure}
\vspace*{0.3em}
 \centerline{\includegraphics[width=8.2truecm]{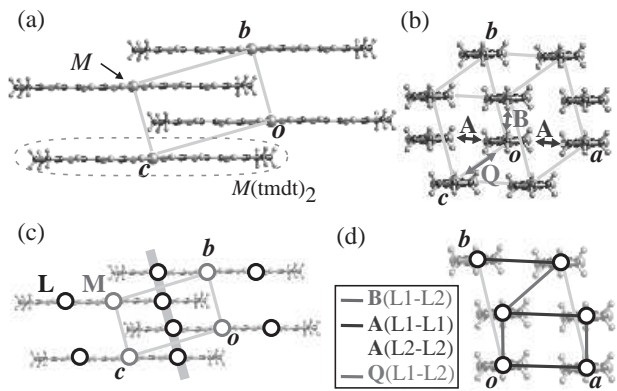}}
\vspace*{-1.8em}
\caption{
Crystal structure of $M$(tmdt)$_2$ ($M$=Ni, Au) [(a) and (b)], 
 and a schematic view of the model discussed in this paper 
 where the lattice sites are represented by circles [(c) and (d)]. 
In (b) and (d), some of the intermolecular and intersite bonds 
 with large transfer integrals (see text) are shown. 
In (c), two-dimensional plane of ligands projected onto the $bc$-plane is represented in light blue, 
whose network within the $ab$-plane is shown in (d). 
}
\vspace*{-1.2em}
\label{fig1}
\end{figure}

The success in synthesizing single-component molecular metals (SCMM)
 by the Kobayashis and co-workers~\cite{Tanaka2001Science,ReviewSCMM1,ReviewSCMM2}  
 opened a new route in realizing molecule-based electronic conductors. 
These compounds intrinsically possess conducting nature 
 by self-assembly of same kind of neutral molecules, 
 in sharp contrast with conventional molecular conductors 
 composed of two or more chemical species with charge transfer (CT) between them, 
 which was the unique way in realizing charge carriers in molecular crystals for a long time. 
The first SCMM synthesized was Ni(tmdt)$_2$,~\cite{Tanaka2001Science}  
 whose crystal structure is shown in Figs.~\ref{fig1}(a) and (b);  
 its electrical resistivity $\rho(T)$ decreases as decreasing temperature 
 down to 0.6~K 
 and the existence of a Fermi surface (FS) was confirmed by de Haas-van Alphen oscillation~\cite{Tanaka2004JACS}. 
Among several SCMM reported since then~\cite{ReviewSCMM1,ReviewSCMM2}, 
 Au(tmdt)$_2$~\cite{Suzuki2003JACS}, 
 which is isostructural to its Ni analog and likewise metallic down to low temperatures~\cite{Tanaka2007CL}, 
 has attracted interest because of its high antiferromagnetic (AF) transition temperature among molecular conductors
 at $T_{\rm AF}$=110~K~\cite{Suzuki2003JACS,Zhou2004JACS,HaraPreprint}. 
In this study we focus on the electronic properties of these two compounds, 
 $M$(tmdt)$_2$ with $M$=Ni and Au, to shed light on the nature of 
 the electronic states in this new class of materials.%

All SCMM known to date are composed of transition metal complex molecules 
 with extended-TTF dithiolene ligands 
 of the form $M$($L$)$_2$ ($M$: metal, $L$: ligand), 
 reflecting the strategy of ``frontier molecular orbital (MO) designing"~\cite{AKobayashi2001JMatChem}. 
They are chosen to satisfy the desired conditions 
 for realizing SCMM: 
 (i) small energy separations $\Delta E$ between frontier MO in the isolated molecules, and 
 (ii) existence of large inter-molecular transfer integrals when crystals are formed. 
As discussed in ref.~\ref{AKobayashi2001JMatChem} in detail, 
 $M$($L$)$_2$ molecules are suitable for condition (i), 
 since their frontier MO are usually characterized by bonding and anti-bonding 
 combinations of ``virtual" frontier MO of the two ligands. 
Then $\Delta E$ is roughly twice the effective transfer integral between them,  
 which can be tuned small by using appropriate $L$ and $M$. 
As for condition (ii), 
 TTF-type skeletons 
 produce large inter-molecular overlaps efficiently 
 as heavily used in CT salts. 
When these conditions are fulfilled for molecules with even number of electrons, 
 widths of the bands near the Fermi energy, $\ef$, 
 originated from frontier MO can exceed $\Delta E$, 
 leading to band overlaps so that FS would exist, i.e., resulting in semi-metallic states.  
As for molecules with odd number of electrons (which are rather rare) 
 the latter condition is necessary to avoid localization due to electron correlation, 
 typically Mott insulating states. 

The band structures of Ni(tmdt)$_2$ and Au(tmdt)$_2$ have been calculated 
 by the extended H{\"u}ckel tight-binding scheme~\cite{Tanaka2001Science}
 as well as by {\it ab-initio} first-principles calculations~\cite{Rovira2002PRB,Ishibashi2005JPSJ,IshibashiPreprint}. 
The results show that several bands originated from different MO 
 are situated near $\ef$ 
 with appreciable mixings between them, 
 consistent with the above-mentioned strategy. 
Note that this is very different from most of the CT-type molecular conductors 
 where the bands near $\ef$ are formed by 
 one $\pi$-character frontier MO per molecule. 
There, effective Hubbard-type models have been successfully applied
 to study their low-energy physical properties theoretically~\cite{SeoReview}. 
The purpose of this work is to investigate whether we can make use of 
 such effective models for SCMM as well. 
In the following, 
 by comparing the first-principles band structures and MO of isolated molecules, 
 we construct a three-band Hubbard model based on virtual MO of the molecules, 
 elaborating the above-mentioned two ligands MO picture~\cite{AKobayashi2001JMatChem}. 
Furthermore, 
 we investigate this model to pursue the origin of the AF transition observed in Au(tmdt)$_2$, 
 which has theoretically been proposed by Ishibashi {\it et al.}~\cite{Ishibashi2005JPSJ} 
 to be due to FS nesting based on first-principles calculations.

\begin{figure}
\vspace*{0em}
 \centerline{\includegraphics[width=8.2truecm]{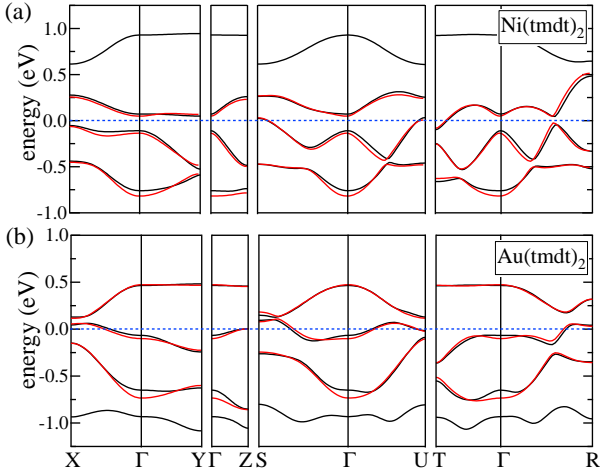}}
\vspace*{-2em}
\caption{
Band structures of (a) Ni(tmdt)$_2$ and (b) Au(tmdt)$_2$ near the Fermi energies, 
 between $\Gamma$(0,0,0)-point, and X($\pi$,0,0), Y(0,$\pi$,0), Z(0,0,$\pi$), 
 S($\pi$,$\pi$,0), U($\pi$,0,$\pi$), T(0,$\pi$,$\pi$), and R($\pi$,$\pi$,$\pi$)-points (black lines). 
Fitted results and the Fermi levels are shown by red solid and blue dotted lines.
}
\vspace*{-1.5em}
\label{fig2}
\end{figure}%
These compounds are nominally written as 
 Ni$^{2+}$(tmdt$^-$)$_2$ and Au$^{3+}$(tmdt$^{1.5-}$)$_2$, 
 with metal ions having (3d)$^8$ and (5d)$^8$ configurations~\cite{ReviewSCMM1,ReviewSCMM2}.  
Since neutral tmdt has even number of electrons, 
 the Ni (Au) salt has even (odd) number of electrons per molecule. 
In Fig.~\ref{fig2}, 
 their band structures near $\ef$
 calculated by first-principles method 
are shown~\cite{Ishibashi2005JPSJ,IshibashiPreprint}.
Due to the resemblances in their electronic configuration and crystal structure, 
 overall features of the four bands shown in the figure are similar between the two compounds.
A difference is seen in 
 the top band which is isolated in Ni(tmdt)$_2$,
 whereas it is close to the next band near the X and S points in Au(tmdt)$_2$.
In contrast, 
 in the case of Ni(tmdt)$_2$ the bottom band almost touches to the third band from the top  
 in several regions of the Brillouin zone, 
 while in Au(tmdt)$_2$ it is separated from the others. 
Therefore three bands each are relevant near $\ef$, 
 i.e., the lower (upper) three bands for the Ni (Au) salt, 
 which we will consider in the following. 
The even/odd electrons per molecule in the two compounds 
 result in electron and hole pockets with equal volume in Ni(tmdt)$_2$ 
 and a half-filled band in Au(tmdt)$_2$.%

\begin{figure}
\vspace*{0em}
 \centerline{\includegraphics[width=8.2truecm]{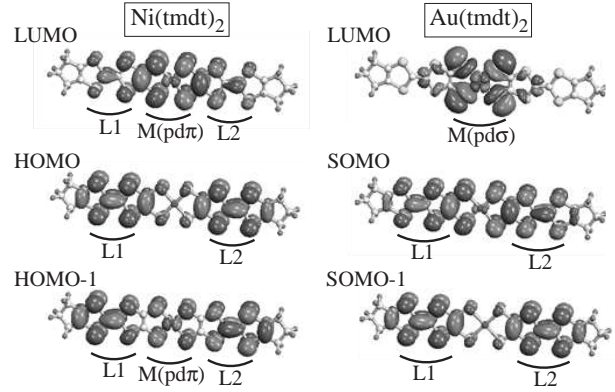}}
\vspace*{-1em}
\caption{
Molecular orbitals from which the bands near the Fermi energy are formed~\cite{IshibashiPreprint}. 
The rough spatial extensions of the L and M orbitals 
 explained in the text are indicated. 
}
\vspace*{-1.7em}
\label{fig3}
\end{figure}

These bands near $\ef$ originate from three MO for each molecule, 
 which are shown in Fig.~\ref{fig3}~\cite{IshibashiPreprint}. 
A close resemblance is seen between HOMO of Ni(tmdt)$_2$ 
 and SOMO-1 of Au(tmdt)$_2$ 
 which, as discussed in ref.~\ref{AKobayashi2001JMatChem}, 
 have the bonding nature between the two ligands MO. 
On the other hand, 
 in the corresponding anti-bonding orbitals, i.e.,  
 LUMO of Ni(tmdt)$_2$ and SOMO of Au(tmdt)$_2$, 
 the contributions from the $d$-orbital at the metal site 
 and the $p$-orbital at the four surrounding S atoms 
 are more prominent in the former than in the latter~\cite{IshibashiPreprint}. 
Considering these characteristics, 
 we attempt to reconstruct MO in Fig.~\ref{fig3} by three parts of the molecules, 
 as follows. 
In Ni(tmdt)$_2$, we can postulate that LUMO, HOMO, and HOMO-1 are approximately described by 
 anti-bonding, non-bonding, and bonding combinations of three virtual MO within the molecule, 
 $\phi_{\rm L1}-\phi_{\rm L2}+c_\textrm{ab}\phi_{\textrm{M}(pd\pi)}$, $\phi_{\rm L1}+\phi_{\rm L2}$, and 
 $\phi_{\rm L1}-\phi_{\rm L2}-c_\textrm{b}\phi_{\textrm{M}(pd\pi)}$, respectively. 
Here, $\phi_{\rm L1}$ and $\phi_{\rm L2}$ are the two ``ligand (L) orbitals" 
 close to LUMO of the embedded TTF molecule, 
 and $\phi_{\textrm{M}(pd\pi)}$ is the ``metal (M) orbital" 
 centered at the Ni site, and $c_\text{ab}$ and $c_\text{b}$ are some coefficients. 
Note that we consider this M orbital 
 not as a bare $3d$ orbital of Ni as considered in ref.~\ref{AKobayashi2001JMatChem}, 
 but rather close to the anti-bonding $pd\pi$ orbital between  
 the $3d_{xz}$-orbital\cite{noteXYZinDorb} of the Ni atom and the four $2p$-orbitals of the surrounding S atoms. 
As for Au(tmdt)$_2$, 
 LUMO is another ``M orbital" roughly being the anti-bonding $pd\sigma$ orbital, 
 $\phi_{\textrm{M}(pd\sigma)}$, 
 between the Au $5d_{xy}$-orbital\cite{noteXYZinDorb} and the surrounding S $2p$-orbitals~\cite{IshibashiPreprint}. 
On the other hand, by neglecting the contribution from $\phi_{\textrm{M}(pd\pi)}$ smaller than in Ni(tmdt)$_2$, 
 SOMO and SOMO-1 of Au(tmdt)$_2$ are described as $\phi_{\rm L1}-\phi_{\rm L2}$ and $\phi_{\rm L1}+\phi_{\rm L2}$. 
We stress here that the occupation probability of the $d$-orbitals itself is small in both M orbitals, 
 but their characters show up through hybridization with the S $2p$-orbitals.%

Then, 
 assuming the above two L and one M orbitals, 
 namely, L1, L2, and M($pd\pi$) for Ni(tmdt)$_2$ or M($pd\sigma$) for Au(tmdt)$_2$, 
 as base orbitals, 
 we can construct an effective three-band Hubbard model for each material 
 written as%
\begin{align}
{\cal H} = &\ t_{ij} \sum_{ijs}  \left( c^\dagger_{is} c_{js}^{} + \mathrm{h.c.} \right)  
 + \Delta \sum_{\ i(\textrm{M})} (n_{i\uparrow} + n_{i\downarrow}) \nonumber\\
 & + U_{\textrm{L}} \sum_{i(\textrm{L})} n_{i\uparrow} n_{i\downarrow} 
 + U_\textrm{M} \sum_{i(\textrm{M})} n_{i\uparrow} n_{i\downarrow}, 
\label{Ham}
\end{align}
where $c^\dagger_{is}$ is the creation operator of an electron 
 with spin $s(= \uparrow\!\! / \!\! \downarrow)$ at the $i$th site, 
 and $n_{is} \equiv c^\dagger_{is} c_{is}^{}$. 
$t_{ij}$ are transfer integrals between site pairs $i$ and $j$, 
 and $U_\textrm{L}$ and $U_\textrm{M}$ are on-site Coulomb energies 
 for the L and M orbitals, respectively. 
The one-particle energy difference between the two kinds of orbital is denoted as 
 $\Delta$, which is measured from the L orbital. 
The summation $\sum_{ijs}$ is taken for all site pairs considered (L-L, L-M, and M-M pairs), 
 while $\sum_{i(\textrm{L})}$ and $\sum_{i(\textrm{M})}$ are for  
 sites belonging to the L and M orbitals, respectively. 

To find out the appropriate parameters in eq.~(\ref{Ham}), 
 we perform numerical fitting of 
 the energy dispersion for its paramagnetic Hartree-Fock solution, 
 in order to maintain consistency with the spin-dependent calculations below,  
 to the first-principle band structures. 
We set $U_\textrm{L}$=$U_\textrm{M}$ $\equiv$ $U$ for simplicity, 
 which are considered to be of the same order deduced 
 from similar spatial extensions of the two orbitals seen in Fig.~\ref{fig1}. 
The fitted results for $U=0.3$~eV are shown in Fig.~\ref{fig2}, 
 which show good agreement with the first-principles band structure, 
 particularly near $\ef$. 

%
\begin{table}
\caption{Fitted values of $t_{ij}$ and $\Delta$ in eq.~(\ref{Ham}) with $U_\textrm{L}$=$U_\textrm{M}$ $\equiv$ $U=$~0.3~eV. 
For example, 
 ``A[100] L1-M" denotes $t_{ij}$ between L1 orbital belonging to $M$(tmdt)$_2$ at the origin 
 and M orbital of $M$(tmdt)$_2$ at ($a$, 0, 0). 
}
 \vspace{0.2em}
 \begin{center}
 \begin{tabular}{ccccc}
 & site pair & Ni(tmdt)$_2$ & Au(tmdt)$_2$ \\ \hline
intra-mol. & L1-L2 & 11.3 meV & -54.3 meV \\ 
A[100] & L1-L1, L2-L2 & {\bf -80.8} & {\bf -95.5} \\
A[100] & L1-L2 & -7.8 & -5.4 \\
B[111] & L1-L2 & {\bf 216.0} & {\bf 208.0} \\
C[101] & L1-L2 & 9.7 & 13.0 \\
Q[001] & L1-L1, L2-L2 & 9.0 & 3.6 \\
Q[001] & L1-L2 & {\bf 123.0} & {\bf 108.7} \\
P[211] & L1-L2 & 39.8 & 37.4 \\
R[011] & L1-L2 & 9.7 & 4.2 \\ \hline
intra-mol. & L1-M, M-L2 & {\bf -218.0} & --- \\ 
A[100] & L1-M, M-L2 & 14.9 & 38.3 \\
A[100] & M-L1, L2-M & 14.3 & 1.0 \\
B[111] & L1-M, M-L2 & 41.9 & 28.0 \\
C[101] & L1-M, M-L2 & -21.1 & -17.9 \\
Q[001] & L1-M, M-L2 & 26.1 & 11.6 \\
R[011] & L1-M, M-L2 & 1.8 & 17.2 \\ \hline
A[100] & M-M & -21.5 & {\bf 94.3} \\
B[111] & M-M & 22.3 & -1.0 \\
C[101] & M-M & -0.6 & 1.8 \\
Q[001] & M-M & -3.3 & 1.5 \\ \hline
$\Delta$ & & 11.5	& 717.7 \\
 \end{tabular}
 \end{center}
\vspace*{-1.5em}
\label{table1}
\end{table}%
The fitted parameters are listed in Table~\ref{table1}. 
$t_{ij}$ for the L-L pairs are all similar between the two compounds, 
 naturally expected since they are isostructural. 
There are several inter-molecular $t_{ij}$ 
 with absolute values much larger than that for the intra-molecular L1-L2 pair, 
 which form a two-dimensional network in the (001) plane as shown in Figs.~\ref{fig1}(c) and (d).
There is a large $t_{ij}$ for the bond along B[111] between the different ligands (L1-L2),  
 of about 0.2 eV which is twice larger than the others; they form dimers. 

The transfer integrals for L-M and M-M pairs show different aspects 
 between the two compounds. 
In Ni(tmdt)$_2$, 
 the intra-molecular L-M pair with $|t_{ij}|\simeq$ 0.22~eV, 
 which is comparable to the above B[111] L1-L2 bond, 
 brigdes the two-dimensional L networks, 
 and make the electronic structure three-dimensional.  
On the other hand, 
 for Au(tmdt)$_2$, 
 the intra-molecular L-M bond 
 is not included in the fitting due to the $\pi$ and $\sigma$ symmetry of the two orbitals; 
$t_{ij}$ for inter-molecular L-M pairs are rather small 
 but responsible for the three-dimensional $\pi$-$pd\sigma$ hybridization. 

The L-M energy difference $\Delta$ in Ni(tmdt)$_2$ is surprisingly small, 
 i.e., the two kinds of orbitals are almost degenerate in energy, 
 indicating that the three bands are formed by completely mixed L and M orbitals. 
This suggests that both orbitals play crucial roles in forming the bands crossing $\ef$, 
 and therefore, in realizing the metallic state in this compound, 
 in contrast with the two ligand MO picture 
 as has been discussed in the litteratures~\cite{AKobayashi2001JMatChem,Rovira2002PRB}.
As for Au(tmdt)$_2$, 
 the fitted value for $\Delta=0.72$~eV suggests that  
 the top band is formed mainly by the M($pd\sigma$) orbital~\cite{Ishibashi2005JPSJ}. 
We note that the difference in the value of $U$ used in the fitting 
 barely affects the parameters for Ni(tmdt)$_2$, 
 but for Au(tmdt)$_2$ $\Delta$ is affected 
 while $t_{ij}$ are not; 
 $\Delta$ should be considered as an effective L-M energy difference.  

\begin{figure}
\vspace*{0.5em}
 \centerline{\includegraphics[width=9truecm]{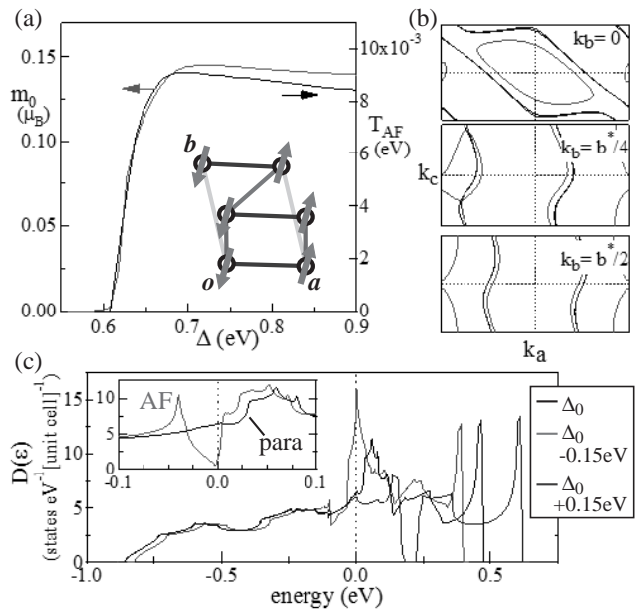}}
\vspace*{-1.5em}
\caption{
(a) Transition temperature $T_\textrm{AF}$ and 
 $T=0$ magnetic moment $m_0$, 
 for the antiferromagnetic (AF) state, 
 (b) Fermi surface (the Brillouin zone is deformed to a rectangular solid), 
 and (c) density of states $D(\epsilon)$, for the paramagnetic metallic state 
 in the Hartree-Fock calculation for Au(tmdt)$_2$ by varying $\Delta$ 
 (the inset shows change in $D(\epsilon)$ near $\ef$ between the paramagnetic and AF states for $\Delta=\Delta_0$). 
A schematic view of the AF ordered spin pattern of the L orbital sites 
in the $ab$-plane is shown in the inset of (a), 
whereas M sites have no moment. 
}
\vspace*{-1em}
\label{fig4}
\end{figure}

Next we will proceed to the analysis of the model above for Au(tmdt)$_2$ 
 and discuss its magnetic instability. 
In the following we investigate the electronic properties of this model 
 by varying $\Delta$ from its fitted value ($\equiv\Delta_0$)
 while fixing other parameters, i.e., 
 $t_{ij}$ as in Table~\ref{table1} and $U$=0.3~eV. 
Within spin-dependent Hartree-Fock approximation, 
 we have sought for magnetically ordered solutions 
 with commensurate AF spin patterns 
 up to 2$\times$2$\times$2 of the original unit cell, 
 and a stable solution consistent with that obtained in ref.~\ref{Ishibashi2005JPSJ} 
 is found. 
This state has anti-parallel magnetic moments on the two L orbitals within each molecule, 
 and characterized by the wave vector ${\bf q}=(a^*/2,0,0)$, 
 whereas no moment appears on the M sites 
 (see Fig.~8 in ref.~\ref{Ishibashi2005JPSJ} and the inset of Fig.~\ref{fig4}~(a)) 
In Fig.~\ref{fig4}(a), 
 the calculated transition temperature $T_\textrm{AF}$ for this solution 
 as well as its magnetic moment at $T=0$, $m_0$, are shown 
 as a function of $\Delta$. 
The magnetically ordered state has small but finite density of state at $\epsilon_\textrm{F}$, 
 $D(\epsilon_\textrm{F})$, as seen in the inset of Fig.~\ref{fig4}~(c), 
 i.e., the system is still metallic~\cite{Ishibashi2005JPSJ}. 
$T_\textrm{AF}$ and $m_0$ show a maximum at around $\Delta=\Delta_0$, 
 whose origin is understood as follows.%

In Fig.~\ref{fig4}~(b) and (c), 
 we show the FS  
 and the density of states $D(\epsilon)$ 
 of the paramagnetic solution,  
 for $\Delta=\Delta_0$ and $\Delta_0 \pm 0.15$ eV. 
One can see that FS changes little when $\Delta$ is increased from $\Delta_0$, 
 and its nesting property found in ref.~\ref{Ishibashi2005JPSJ} is maintained; 
 therefore the nesting is due to the $\pi$-network of L orbital. 
When decreasing $\Delta$ from large values, 
 the slight increase of $T_\textrm{AF}$ and $m_0$ at $\Delta \gsim \Delta_0$ 
 is due to the enhancement of $D(\epsilon_\textrm{F})$ 
 by the increase of L-M hybridization as seen in Fig.~\ref{fig4}~(c). 
This gives rise to an increase of the whole profile of the bare susceptibility 
 $\chi_0(\textbf{q})$ (not shown; see Fig.~6 in ref.~\ref{Ishibashi2005JPSJ}), 
 directly related to the magnetic instability. 
Further decrease of $\Delta$ actually raises $D(\epsilon_\textrm{F})$ more rapidly, 
 but leads to modification of FS and easily destroys the nesting,  
 which results in the sudden decrease in $T_\textrm{AF}$ and $m_0$. 
We note that we find a shift of peak position in $\chi_0({\bf q})$ to an incommensurate wave vector 
 $\textbf{q}=(a^*/2,\delta,0)$ 
 in a narrow region of $0.59$~eV $\leq \Delta \leq 0.63$~eV,  
 then an incommensurate AF could be stabilized there  
 which is not considered in our calculations. 
Still we expect that the rapid decrease of $T_\textrm{AF}$ holds 
 even when such possibility is included.%

Let us compare our results with the experiments on Au(tmdt)$_2$. 
If the transition at 110 K seen in the experiments is 
 due to the spin-density-wave (SDW) formation by FS nesting, 
 a signature in $\rho (T)$ should appear due to the change in the band structure, 
 as a hump in $\rho (T)$ below the transition temperature as found in metallic Cr~\cite{Fawcett1988RMP}. 
This was not seen in previous measurements on compressed pellet samples~\cite{Suzuki2003JACS} 
 nor on microcrystals~\cite{Tanaka2007CL}, 
 while a small kink has recently been detected by a preliminary measurement on a single crystal~\cite{CuiUnpub}~;
 further measurements on single crystal samples are awaited. 
A possible origin of the high $T_\textrm{AF}$ among molecular conductors 
 is the three-dimensionality of the system as has been discussed in ref.~\ref{HaraPreprint}. 
This is not realized in conventional CT salts showing SDW, e.g., in TMTSF$_2X$, 
 due to the existence of 
 the anions or cations providing carriers into the $\pi$ network 
 resulting in low-dimensional electronic structures, 
 which lower transition temperatures in general due to fluctuation effects.
Our spin-dependent Hartree-Fock results, 
 which can be interpreted as such SDW state, 
 actually show the same order of $T_{\rm AF}$ to the experiments. 

However, 
 as for the magnetic moments in the ordered state, 
 analysis of a recent $^1$H-NMR experiment suggested that they are rather large, 
 more than 0.6~$\mu_{\rm B}$ per tmdt, 
 inconsistent with our results showing saturated moments at $T=0$ 
 less than 0.2 $\mu_{\rm B}$ per tmdt.
Such large magnetic moments in molecular conductors are reminiscent of 
 strongly correlated systems such as 
 Mott insulating states in the $\kappa$-ET$_2X$ family~\cite{Kanoda2006JPSJ}
 or charge ordered states in the TMTTF$_2X$ and $\theta$-ET$_2X$ families~\cite{Takahashi2006JPSJ}. 
In fact, 
 the AF ordered spin pattern we disscussed in this paper as shown in the inset of Fig.~\ref{fig4}(a) 
 can also be considered as ``AF between dimers"\cite{Kino1996JPSJ}, 
 i.e., ferromagnetic within dimers formed by the B[111] L1-L2 pairs and AF between them, 
 and an insulating gap opens when larger $U$ ($\gsim$~0.4 eV) is used in the calculation. 
This state can be viewed as the dimer-Mott insulator~\cite{Kino1996JPSJ,SeoReview}, 
 although it cannot be identified within Hartree-Fock approximation. 
In the above-mentioned strongly correlated states, 
 experimentally the systems show insulating behavior at low temperatures 
 and $\rho (T)$ shows no anomaly at the magnetic transition temperatures; 
 these are both distinct from the experimental data for Au(tmdt)$_2$ 
 with metallic behavior and the kink at $T_{\rm AF}$ in $\rho (T)$.
Nevertheless, the high $T_{\rm AF}$ and large magnetic moments seen in the experiments 
 can be due to the closeness to such dimer-Mott insulating states, 
 whose investigations are left for future studies. 
Finally, 
 we note that a low transition temperature below 10~K was recently found in Au(tmstfdt)$_2$~\cite{FujiwaraUnpublished}, 
 a selenide analog of Au(tmdt)$_2$, 
 despite they are isostructural and have similar electronic structures. 
This might be a result of the fragileness of the FS nesting
 by L-M hybridization~\cite{IshibashiPreprint}, seen in our calculation.

In summary, 
 we have constructed an effective three-band Hubbard model 
 for single-component molecular metals $M$(tmdt)$_2$ ($M$=Ni, Au) 
 which indicates that this system is a multiband $\pi$-$pd\pi$/$pd\sigma$ system. 
We have found that the close degeneracy of $\pi$ and $pd\pi$ ``virtual" MO states 
 in Ni(tmdt)$_2$ is a key to realizing its metallic state. 
The origin of the antiferromagnetic transition observed in 
 Au(tmdt)$_2$ is discussed based on spin-dependent Hartree-Fock calculations. 
We believe that this new category of materials will be 
 promising for finding richer phenomena in molecular crystals. 

\section*{Acknowledgment}
The authors thank K. Kanoda, K. Miyagawa, M. Mochizuki, and H. Tanaka for fruitful discussions. 
This work was supported by a Grant-in-Aid for Scientific Research 
(Nos.~18740221, 18028026, and 19014020)
  from the 
Ministry of Education, Culture, Sports, Science and Technology.


\begin{thebibliography}{99}
\bibitem{noteXYZinDorb}
 The axes are chosen as $x$ and $y$ along the long and short axes within the molecular plane 
 and $z$ perpendicular to it.
\end{thebibliography}
\end{document}